\documentclass[fleqn,usenatbib]{mnras}

\usepackage{newtxtext,newtxmath}

\usepackage[T1]{fontenc}

\DeclareRobustCommand{\VAN}[3]{#2}
\let\VANthebibliography\thebibliography
\def\thebibliography{\DeclareRobustCommand{\VAN}[3]{##3}\VANthebibliography}

\usepackage{graphicx}	%
\usepackage{amsmath}	%

\newcommand{\be}{\begin{equation}}
\newcommand{\ee}{\end{equation}}
\newcommand{\beqa}{\begin{eqnarray}}
\newcommand{\eeqa}{\end{eqnarray}}

\title[SMBHs from mergers and accretion in NSCs]{Supermassive black holes from runaway mergers and accretion in nuclear star clusters}

\author[Kritos, Berti, Silk]{
Konstantinos Kritos,$^{1}$\thanks{E-mail: kkritos1@jhu.edu}
Emanuele Berti,$^{1}$
and Joseph Silk$^{1,2,3}$
\\
$^{1}$William H. Miller III Department of Physics and Astronomy, Johns Hopkins University, Baltimore, Maryland 21218, USA\\
$^{2}$Insitut d'Astrophysique de Paris, UMR 7095 CNRS and UPMC, Sorbonne Universit\'{e}, F-75014 Paris, France\\
$^{3}$Department of Physics, University of Oxford, Oxford OX1 3RH, United Kingdom
}

\date{Accepted XXX. Received YYY; in original form ZZZ}

\pubyear{2024}

\begin{document}
\label{firstpage}
\pagerange{\pageref{firstpage}--\pageref{lastpage}}
\maketitle

\begin{abstract}
Rapid formation of supermassive black holes occurs in dense nuclear star clusters that are initially gas-dominated. Stellar-mass black hole remnants of the most massive cluster stars sink into the core, where a massive runaway black hole forms as a consequence of combined effects of repeated mergers and Eddington-limited gas accretion. The associated gravitational wave signals of high-redshift extreme mass-ratio inspirals are a unique signature of the nuclear star cluster scenario.
\end{abstract}


\noindent {\bf \em Introduction.}
The emergence of supermassive black holes (SMBHs) in the early Universe is an outstanding problem in modern cosmology and astrophysics.
Many active galactic nuclei (AGN) powered by SMBHs have been identified at redshift $z \gtrsim 6$, with spectroscopically inferred black hole (BH) masses in the range
$10^{6}$--$10^8 \,M_\odot$~\citep {2024ApJ...964...39G,2024arXiv240303872J,Maiolino:2023zdu}.
As suggested by recent spectroscopic surveys with the James Webb Space Telescope (JWST), the abundance of red-selected AGN is a hundred times that expected in this redshift range~\citep{Matthee:2023utn,2023arXiv230801230M}.
The consensus is that SMBHs have grown out of massive seed black holes~\citep{2012Sci...337..544V},
and the problem is then reduced to explaining the formation of those intermediate-mass BH seeds~\citep{2020ARA&A..58..257G}.
One possible formation channel is the direct collapse of pristine gas clouds at $z>10$~\citep{Begelman:2006db,Latif:2022vwc}.
Alternative seeding mechanisms involve repeated BH mergers in dense star clusters~\citep{1987ApJ...321..199Q,1989ApJ...343..725Q,2011ApJ...740L..42D,Lupi:2014vza,Antonini:2018auk,2020MNRAS.498.5652K,Fragione:2020nib,Atallah:2022toy,Kritos:2022non} and feeding by debris of successive tidal disruption events~\citep[TDEs;][]{Stone:2016ryd,Inayoshi:2019fun,Rizzuto:2022fdp}.
Moreover, several works have studied the rapid formation of supermassive stellar objects with masses in the range $10^3$--$10^5M_\odot$ within metal-poor massive star clusters~\citep{2018MNRAS.476..366B,2020ApJ...892...36T,2021MNRAS.503.1051D,2022MNRAS.512.6192S,2023MNRAS.521.3553R} and nuclear star clusters~\citep{2021ApJ...908...57E,2023MNRAS.522.4224V} through runaway collisions of stars and gas accretion. These supermassive stellar objects can then collapse into intermediate-mass BH seeds.

Nuclear star clusters (NSCs) are extremely dense and massive stellar systems that have been observed to reside in the centers of most galaxies in the local Universe~\citep{2020A&ARv..28....4N}.
In the nearby Universe, NSCs span the mass range from $10^5M_\odot$ to $10^9M_\odot$, with evidence for in situ star formation in the more massive systems~\citep{2022A&A...658A.172F}.
At high redshift, a number of lensed star clusters with effective radii as small as a parsec ($\rm pc$) and with masses of order $10^6M_\odot$ have been identified~\citep{2023ApJ...945...53V,2024arXiv240103224A}. There is some theoretical support for the fragmentation of the central region of protogalaxies into millions of metal-poor stars, forming clusters with half-mass radii $\lesssim 0.5\,\rm pc$~\citep{2010MNRAS.409.1057D}.

In this paper, we use a semianalytic NSC model and demonstrate the possibility of rapidly forming SMBHs through the combined processes of repeated BH mergers and Eddington-limited gas accretion, assuming that a substantial amount of gas can be retained within the cluster for a few tens of millions of years ($\rm Myr$).

\noindent {\bf \em Nuclear star cluster model.}
Our numerical model for the NSC implements a two-mass system consisting of $N_\star$ stars, each of mass $m_\star$, and $N_{\rm BH}$ stellar BHs, each of mass $m_{\rm BH}$ and nonspinning, with an initial mass $M_{\rm g}$ of residual gas not turned into stars.
We find that typically one of the centrally located and initially stellar mass BHs grows dramatically via gas, star, and BH accretion, and we denote its mass and spin by $M_{\rm BH}$ and $\chi_{\rm BH}$, respectively.
The total cluster mass is therefore $M_{\rm cl}=m_\star N_\star + m_{\rm BH}N_{\rm BH} + M_{\rm g} + M_{\rm BH}$.
Since the growing BH is one of the stellar BHs, initially $M_{\rm BH,0}=m_{\rm BH}$ and $\chi_{\rm BH,0}=0$.
This ``oligarchic'' growth scenario can occur in dynamical environments \citep[see e.g.][]{Matsubayashi:2004bd,Kovetz:2018vly}, where the formation of multiple massive BHs within the same system is significantly suppressed by their preferential accretion by the most massive object~\citep{Kritos:2022non}.

The gas, composed primarily of ionized hydrogen with sound speed $c_{\rm s}\simeq10\,\rm km\, s^{-1}$~\citep{2009ApJ...703.1352K}, is removed exponentially over a characteristic gas expulsion timescale $\tau_{\rm ge}$~\citep{Baumgardt:2007av}.
The growing BH may accrete some of this gas. We assume an Eddington-limited Bondi accretion rate onto the BH seed $\dot{M}_{\rm acc}$ with Salpeter timescale of $50\,\rm Myr$, and a time-evolving Plummer gas density $\rho_{\rm g}=3M_{\rm g}/[4\pi (r_{\rm h}/1.3)^3]$, where $r_{\rm h}$ is the half-mass radius.
As a consequence of gradual gas removal, $r_{\rm h}$ expands adiabatically~\citep{1980ApJ...235..986H}.

The half-mass relaxation time $\tau_{\rm rh}$, modified for two-mass systems following Eq.~(10) from \cite{Antonini:2019ulv}, is the characteristic time over which a fraction $\zeta\simeq0.1$ of the total cluster energy $E_{\rm cl}=-\kappa GM_{\rm cl}^2/(2r_{\rm h})$~\citep[with $\kappa\simeq0.4$, see][]{1971ApJ...164..399S} can be transferred throughout the cluster via two-body relaxation~\citep{Breen:2013vla}.
The core is dominated by BHs, which partially decouple from stars due to the impossibility of energy equipartition (``Spitzer instability''). The core collapses on a time-scale $\tau_{\rm cc}$ given by Eq.~(5) from \cite{PortegiesZwart:2002iks}.

At this point, three-body hard binaries (3bb) can form and heat the system.
The rate of energy production $\dot{E}_{\rm cl}=-\zeta E_{\rm cl}/\tau_{\rm rh}$ depends on $r_{\rm h}$, and is controlled by the efficiency of two-body relaxation (``H\'enon's principle'').
We compute the energy generated by a single 3bb $Q_{\rm 3bb}$, and use this  condition to calculate the 3bb rate $R_{\rm 3bb}=\dot{E}_{\rm cl}/Q_{\rm 3bb}$ \citep[as in][page~244]{2003gmbp.book.....H}.
If all BHs but one have been ejected from the system, we assume that energy generation proceeds via tidal disruption of stars accreting onto the runaway BH.
The TDE rate is $R_{\rm TDE}=\dot{E}_{\rm cl}/Q_{\rm TDE}$, where $Q_{\rm TDE}=GM_{\rm BH}\overline{m}_\star/(2r_{\rm T})$ is the heat released from a single TDE and $r_{\rm T}$ is the tidal radius~\citep{Rees:1988bf}.
The main advantage of our method, as in \cite{Antonini:2019ulv}, is that it is not necessary to calculate the properties of the core, while we can incorporate the effect of a cusp around the SMBH.

The differential equations that govern the evolution of the system are given by
\begin{subequations}
\label{eq:differential_system}
    \begin{align}
        {d\overline{m}_\star\over dt}&=-\nu{\overline{m}_\star\over t}\Theta(t-t_{\rm se})\,, \label{eq:dmstar_dt}\\
        {dN_\star\over dt}&=-\xi_{\rm e}{N_\star\over\tau_{\rm rh}} - R_{\rm TDE}\,, \label{eq:dN_stardt} \\
        {dN_{\rm BH}\over dt} &= -k_{\rm ejs}R_{\rm 3bb}\,, \label{eq:dN_BHdt} \\
        {dM_{\rm g}\over dt}&=-{M_{\rm g}\over \tau_{\rm ge}} - \dot{M}_{\rm acc}\,, \\
        {dM_{\rm BH}\over dt}&=\dot{M}_{\rm acc} + f_{\rm TDE}\overline{m}_\star R_{\rm TDE}\,, \\
        {dr_{\rm h}\over dt} &= \left({\zeta r_{\rm h}\over\tau_{\rm rh}} + {2r_{\rm h}\over M_{\rm cl}}{dM_{\rm cl}\over dt}\right)\Theta(t-\tau_{\rm cc}) \nonumber \\&- {r_{\rm h}\over M_{\rm cl}}\left( {N_\star}{d\overline{m}_\star\over dt}  -  {M_{\rm g}\over \tau_{\rm ge}} \right) \,. \label{eq:dr_hdt}
    \end{align}    
\end{subequations}%
Equations~\eqref{eq:dmstar_dt} and~\eqref{eq:dN_stardt} can be compared with \cite{2014MNRAS.442.1265A}. The main difference is a loss term due to TDEs, $R_{\rm TDE}$, on the right-hand side of Eq.~\eqref{eq:dN_stardt}. The first three terms on the right-hand side of Eq.~\eqref{eq:dr_hdt} follow \cite{Antonini:2019ulv}, but we also include an adiabatic expansion term due to gas expulsion. 
The quantity $k_{\rm ejs}$ in Eq.~\eqref{eq:dN_BHdt} is the number of single BHs ejected per unit binary. We set $\nu=0.07$, $t_{\rm se}=2\,\rm Myr$~\citep{Antonini:2019ulv}, and $\xi_{\rm e}=0.0074$~\citep{2014MNRAS.442.1265A}.

At each iteration, we sample $k_{\rm 3bb}$, the number of 3bbs formed, from a Poisson distribution with parameter $\lambda=R_{\rm 3bb}dt$, where $dt$ is the time-step of the simulation.
Every 3bb is formed with components $M_{\rm BH}$ and $m_{\rm BH}$, with its initial semimajor axis $a_0$ determined by sampling the initial hardness parameter $\eta_0\ge1$~\citep{Morscher:2014doa}.
Binaries harden until they reach their ejection ($a_{\rm ejb}$) or gravitational-wave (GW) radius ($a_{\rm GW}$), assuming thermal eccentricity.
To compute the merger remnant mass $m_{\rm r}$, spin $\chi_{\rm r}$, and relativistic recoil---or GW kick $v_{\rm k}$---we implement the equations in Sec.~V.A-C of \cite{Gerosa:2016sys} and references therein, which are based on numerical relativity fits and assume isotropic spin directions.
If $a_{\rm GW}>a_{\rm ejb}$ the binary merges within the cluster through the emission of GWs, and if $v_{\rm k}<v_{\rm esc},$ the merger remnant is retained in the cluster.
In this case,   we discretely update the runaway BH mass, spin, and number by the substitutions $M_{\rm BH}\to m_{\rm r}$, $\chi_{\rm BH}\to \chi_{\rm r}$, and $N_{\rm BH}\to N_{\rm BH}-1$, while in any case of ejection, $M_{\rm BH}\to m_{\rm BH}$, $\chi_{\rm BH}\to0$, and $N_{\rm BH}\to N_{\rm BH}-2$.

\begin{figure}
    \centering
    \includegraphics[width=0.49\textwidth]{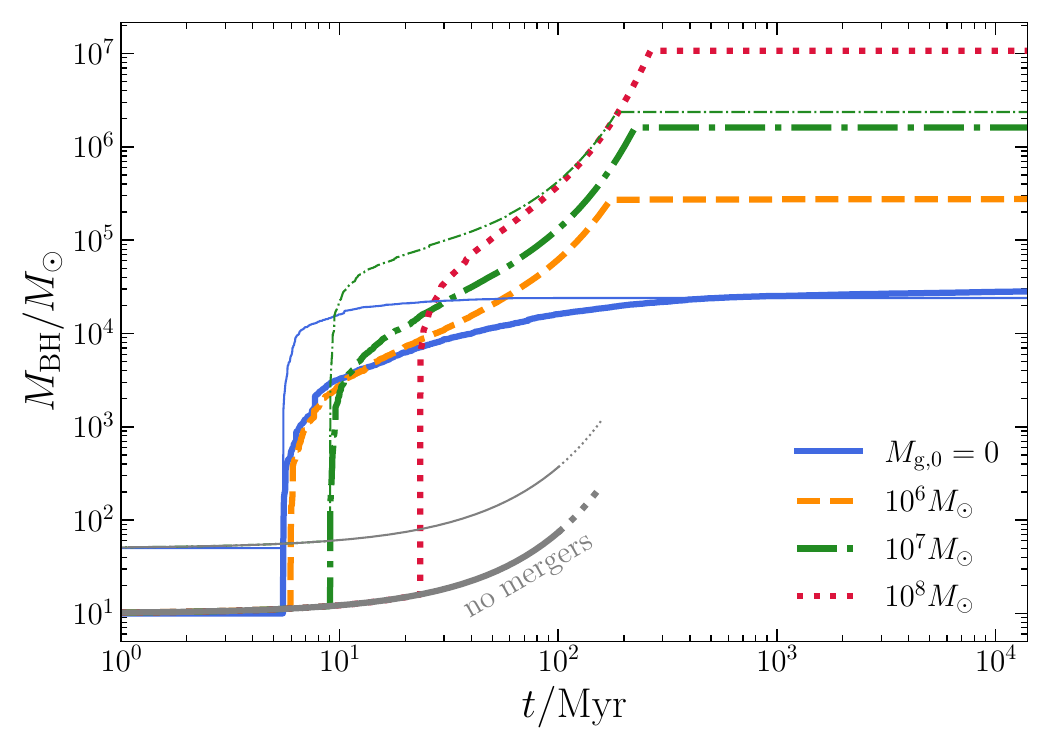}
    \caption{Time evolution of the growing BH mass $M_{\rm BH}$ for different initial amounts of gas in the system. The gas expulsion time-scale is set to $\tau_{\rm ge}=100\,\rm Myr$, the initial stellar number to $N_{\star,0}=10^7$, and the initial half-mass radius to $r_{\rm h,0}=0.1\,\rm pc$. Thick (thin) lines correspond to simulations with $m_{\rm BH}=10M_\odot$ ($m_{\rm BH}=50M_\odot$). For comparison, we also show how a $10M_\odot$ seed (thick gray) and a $50M_\odot$ seed (thin gray) would grow under the sole effect of Eddington accretion.
   }
    \label{fig:black_hole_mass_evolution}
\end{figure}

\noindent {\bf \em Black hole growth.}
We implement a fourth-order Runge-Kutta scheme to numerically solve the system of equations~\eqref{eq:differential_system} using a time step of $dt=0.01\,\rm Myr$ (chosen to be much smaller than $\tau_{\rm rh}$).
We set $N_{\star,0}=10^7$, $r_{\rm h,0}=0.1\,\rm pc$ (where the ``0'' refers to initial values) and $\tau_{\rm ge}=100\,\rm Myr$. Using a Kroupa initial mass function in the range $[0.08,150]M_\odot$, the average initial stellar mass is $\overline{m}_{\star,0}=0.59M_\odot$, while the fraction of BH progenitors with mass $>20M_\odot$ is $\simeq0.18\%$.
We assume $m_{\rm BH}=10M_\odot$, but to quantify uncertainties, we also repeat the analysis with $m_{\rm BH}=50M_\odot$.
In Fig.~\ref{fig:black_hole_mass_evolution}, we show the growth of $M_{\rm BH}$ for four selected values of $M_{\rm g,0}=0,\,10^6,\,10^7,\,10^8M_\odot$.
\begin{figure*}
    \centering
    \includegraphics[width=0.49\textwidth]{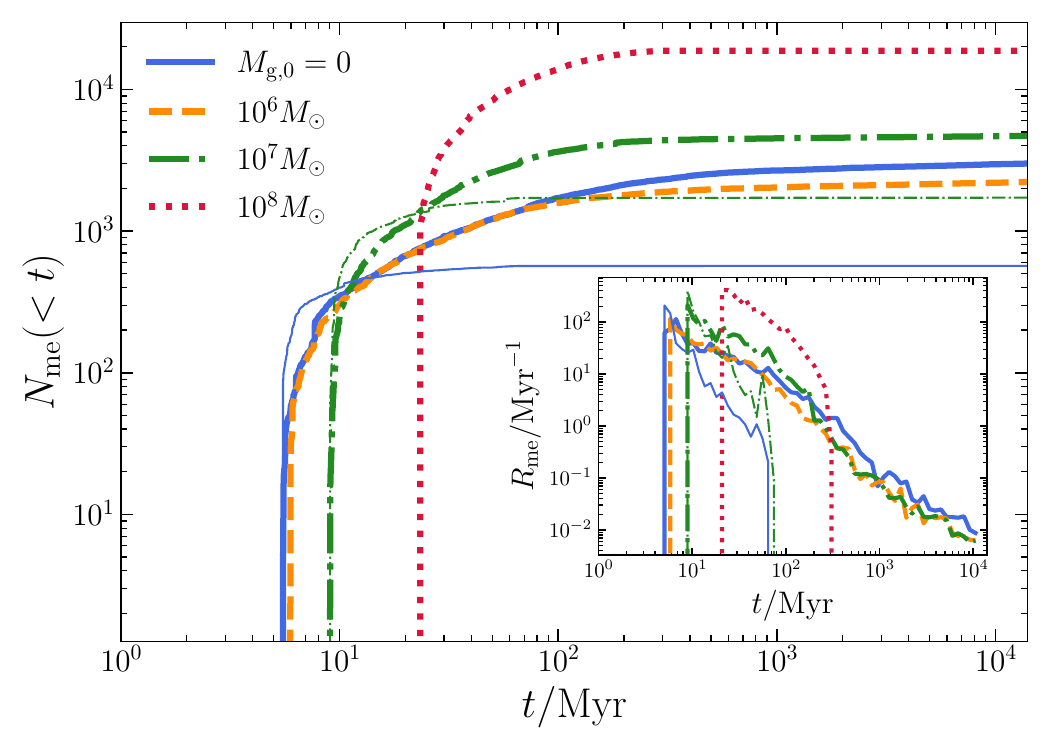}
    \includegraphics[width=0.49\textwidth]{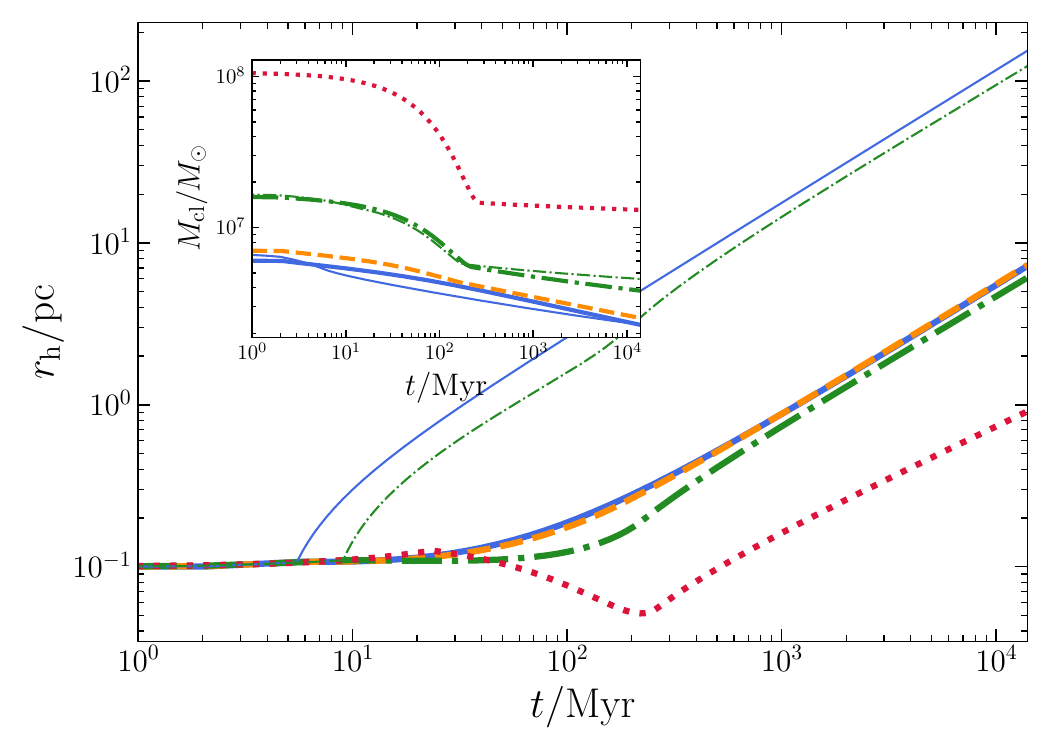}
    \caption{Time evolution of the cumulative number of BH mergers $N_{\rm me}$ (left panel), merger rate $R_{\rm me}$ (left panel, inset), half-mass radius $r_{\rm h}$ (right panel), and cluster mass $M_{\rm cl}$ (right panel, inset) under the same initial conditions used in Fig.~\ref{fig:black_hole_mass_evolution}.}
    \label{fig:merger_rate_global_evolution}
\end{figure*}

\begin{figure*}
    \centering
    \includegraphics[width=\textwidth]{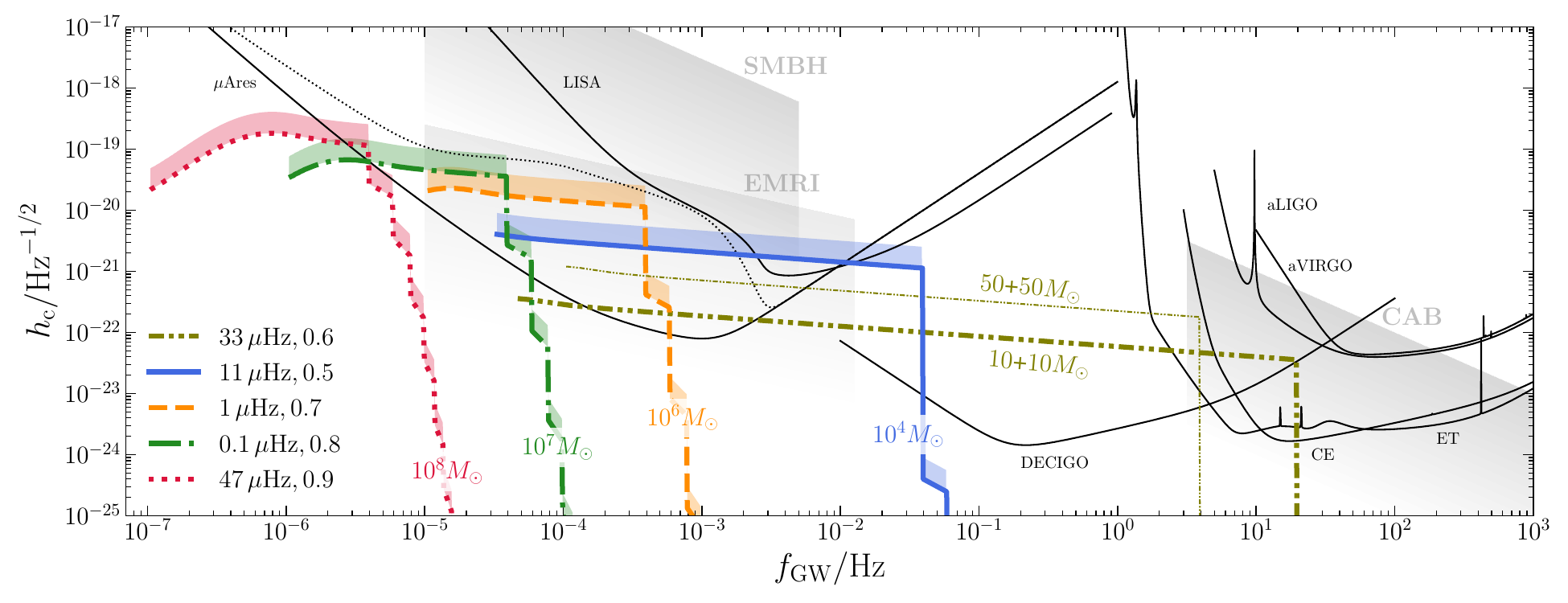}
    \caption{Characteristic GW strain amplitude $h_{\rm c}$ \citep[computed as in][]{Bonetti:2020jku} as a function of detector-frame GW frequency for five extreme mass-ratio inspirals (EMRIs) at redshift $z=10$ (colored lines). We use cosmological parameters from Planck 18~\citep{Planck:2018vyg}. The initial orbital frequency and eccentricity are indicated in the legend; labels on the colored curves refer to the mass of the primary. The bands indicate the $h_{\rm c}$ range as the secondary mass is varied between $10M_\odot$ (lower bound) and $50M_\odot$ (upper bound). The black solid curves correspond to the noise amplitudes of present and planned GW observatories: aLIGO~(\url{https://dcc.ligo.org/LIGO-T1800044/public}), aVIRGO~(\url{https://git.ligo.org/sensitivity-curves/observing-scenario-paper-2019-2020-update}), ET~(\url{https://www.et-gw.eu/index.php/etsensitivities}), CE~(\url{https://cosmicexplorer.org/sensitivity.html}), DECIGO~\citep{Yagi:2011wg}, LISA~\citep{Robson:2018ifk}, and $\mu$Ares~\citep{Sesana:2019vho}. Finally, the shaded regions correspond to lower redshift ($z=1$--$2$) binary sources from stellar-mass compact astrophysical binaries (CAB), EMRIs, and SMBH binaries.}
    \label{fig:gravitational_wave_signal}
\end{figure*}
All simulated systems have $v_{\rm esc,0}>300\,\rm km\, s^{-1}$, allowing repeated mergers to form a runaway BH~\citep{Antonini:2018auk}.
The BH rapidly grows by  hundreds of  repeated BH mergers
(Fig.~\ref{fig:merger_rate_global_evolution}, left panel)
with a merger rate that peaks at $\sim100\,{\rm Myr}^{-1}$ right after core collapse (see inset in the left panel of Fig.~\ref{fig:merger_rate_global_evolution}) and steadily drops as the system expands.

Significant amounts (20\%--30\%) of gas can be accreted within $200\,\rm Myr$, rapidly bringing the BH mass into the SMBH regime.
The residual gas does not fragment because of heating by  UV radiation from cluster stars. Our accretion channel is similar to the direct collapse scenario but occurs on the Salpeter timescale, rather than on the dynamical timescale.
The combination of mergers and accretion helps accelerate the growth because Eddington-limited accretion cannot itself grow a stellar-mass BH seed above $1000M_\odot$ on a similar timescale (compare the gray lines in Fig.~\ref{fig:black_hole_mass_evolution}, which ignore  BH mergers).

Starting from the initial radius $r_{\rm h,0}=0.1\,\rm pc$, the cluster inflates by a factor of 10--100 (depending on $M_{\rm g,0}$), so that $r_{\rm h,0}$ is on the parsec scale after a Hubble time (see Fig.~\ref{fig:merger_rate_global_evolution}, right panel).
When $m_{\rm BH}=50M_\odot$, a larger amount of energy is produced, causing $r_{\rm h}$ to expand to above $100\,\rm pc$ (a feature that resembles the bulges of present-day dwarf galaxies).
The rapid mass loss in the first few hundred Myr is due to gas expulsion (see the inset of the right panel of Fig.~\ref{fig:merger_rate_global_evolution}).
After core collapse, the gas-richest system (dotted red line) undergoes a phase of contraction due to balanced evolution that starts at $\simeq22\,\rm Myr$.
Once the gas has been expelled, the energy production term in Eq.~\eqref{eq:dr_hdt} dominates, and the system starts to expand again at $t\simeq200\,\rm Myr$.

We have demonstrated with an example that a combination of gas accretion and BH mergers is capable of growing stellar-mass BH seeds into the SMBH regime within only $200\,\rm Myr$.
Most notably, if such a cluster forms at $z=15$, then (depending on the initial gas content) an SMBH can assemble by $z\simeq10$.
These SMBH seeds can then further grow by subsequent gas accretion episodes during gas inflows into the NSC following a galaxy merger. This hierarchical scenario can be studied by combining our formalism with galaxy merger trees.

We recall that very massive NSCs are observed locally in many galaxies. 
Many of them are inferred to have gas because we observe young stars. We find that NSC evolution inevitably results in expansion by an order of magnitude or more (see the right panel of Fig.~\ref{fig:merger_rate_global_evolution}). 
In contrast, at formation, many NSCs were massive and compact. This supports our contention that NSCs are a promising environment for generating SMBHs via 
an enhanced merger rate density in the high-redshift Universe.

The ultimate test of our model would be the observation of individual GW signals from SMBH--BH mergers in  NSC cores, which may be possible with future GW observatories (Fig.~\ref{fig:gravitational_wave_signal}).
The GW background produced by these sources is a distinctive signature of the NSC scenario that deserves further study.
Our predictions motivate the development of the science case for next-generation GW observatories spanning the $\mu{\rm Hz}$ frequency range between LISA and Pulsar Timing Arrays.
Our proposed GW background should be taken into account when estimating astrophysical confusion noise sources that can affect the sensitivity of proposed $\mu$Hz detectors, such as $\mu$Ares. For comparison, the black-dotted line in Fig.~\ref{fig:gravitational_wave_signal} shows an estimate of the confusion noise from SMBH and white dwarf binaries in the $\mu$Ares band~\citep{Sesana:2019vho}.

\noindent{\bf \em Acknowledgements.}
We thank Fabio Antonini, Mark~Ho-Yeuk~Cheung, and Giada~Caneva~Santoro for discussions.
This paper was supported by the Onassis Foundation - Scholarship ID: F ZT 041-1/2023-2024.
K.K. and E.B. are supported by NSF Grants No. AST-2006538, PHY-2207502, PHY-090003 and PHY-20043, by NASA Grants No. 20-LPS20-0011 and 21-ATP21-0010, by the John Templeton Foundation Grant 62840, by the Simons Foundation, and by the Italian Ministry of Foreign Affairs and International Cooperation Grant No.~PGR01167.
This work was carried out at the Advanced Research Computing at Hopkins (ARCH) core facility (\url{rockfish.jhu.edu}), which is supported by the NSF Grant No.~OAC-1920103.

\noindent {\bf \em Data availability.}
The output of the simulations is available upon a reasonable request to the corresponding
author. A {\sc Python} implementation of our model through the code {\sc Nuce} (for ``NUclear Cluster Evolution'') is publicly available on {\sc GitHub} at the URL~\url{https://github.com/Kkritos/Nuce}.

\bibliographystyle{mnras}
\bibliography{SMBHformation} %

\bsp	%
\label{lastpage}
\end{document}